\documentclass[12pt]{article}
\usepackage[dvips]{graphicx}
\newcommand{\be}{\begin{eqnarray} }
\newcommand{\ee}{\end{eqnarray} }
\newcommand{\beq}{\begin{equation} }
\newcommand{\eeq}{\end{equation} }

\begin{document}
\begin{center}{\Large \bf Possibility of the new type phase transition}\end{center}
        \begin{center}{\large A.N.~Sissakian,
                O.Yu.~Shevchenko, V.N.~Samoilov}\end{center}
\begin{center}\it
Joint Institute for Nuclear Research,
Dubna, Moscow region 141980, Russia
\end{center}
\begin{abstract}
The dynamical quantum effects arising due to the boundary presence 
with two types of boundary conditions (BC) satisfied by scalar fields are studied. 
It is shown that while the Neumann BC lead to the usual scalar
field mass generation, the Dirichlet BC give rise to the dynamical
mechanism of spontaneous symmetry breaking.
Due to the later, there arises the possibility of the new type phase transition 
from the normal phase to the spontaneously broken one. 
In  particular, at the critical value of the combined evolution parameter the usual massless
scalar QED transforms to the Higgs model.
\end{abstract}
%\pacs{11.10.Wx, 11.30.Qc  }
%\maketitle
The investigation of the quantum field theory (QFT) systems with respect to
their response to the different external influences
(like the different external fields,
nonzero temperature and density of the medium, etc.) 
allows one to discover some new properties of these systems.
For example, it is of  interest to study the phase transitions 
in the QFT systems with the spontaneous symmetry breaking (like
the Higgs model \cite{1}) at nonzero temperature \cite{2,3}.
It is of importance that the temperature (just as well as the finite medium 
density) always restores
the initially broken symmetry and the phase transition from the broken to the 
normal (unbroken) phase 
occurs with the temperature
increase \cite{2,3}.

On the other hand, it is possible to arrive at the very interesting class of external influences
if one considers the QFT system quantized not in the infinite space, as usual, but  
in the space restricted by some boundary surfaces with the 
respective BC 
satisfied by the fields.
Such situations arise in physics very often. These are, for example:  
potential barriers for scalar mesons modeled by the Dirichlet and (or) 
Neumann BC 
in nuclear physics;
the Casimir BC satisfied by the electromagnetic field on metal surfaces in QED, the impenetrable
for the quarks and gluons nucleon surface modeled by the bag BC in QCD. It is well known that
the Casimir effect occurs in all these cases (see \cite{4} for the excellent review). 
However, the Casimir effect is the effect of zero order 
in the coupling constant and deals with the free 
fields\footnote{The corrections to the Casimir force
and energy due to the interaction were considered in \cite{6}.
}. So, it is of interest to consider           
the possibilities of some purely dynamical (caused by interaction) phenomenons in the 
boundary presence. In particular, we will be especially interested in the possibility of the
dynamical (and depending on the characteristic region size) particle mass generation 
in the initially massless theories. Namely such situation occurs in QFT at
finite temperature, for example in the  scalar field theory \cite{3}, where 
the initially massless particle becomes massive due to the temperature 
inclusion while the nontrivial (depending on the temperature) part of dynamical mass disappears
in the zero temperature limit.    

Let us consider the {\it massless scalar field theory} with 
${\cal L}_{int}=\lambda\varphi^4/4!$
quantized in the flat gap pictured on Fig. 1.
\begin{figure}[ht!]
\begin{center}
\includegraphics[height=3.8cm,width=5.3cm]{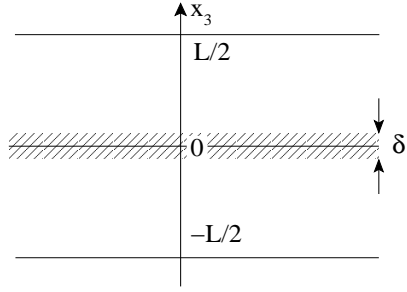}
\end{center}
\caption{The flat gap with the shadowed central $\delta$-region.}
\label{fig1}
\end{figure}
We will consider two possible types of 
BC satisfied by the field $\varphi$ on the plates. These are 
Dirichlet BC:
\be
\label{eq1}
\varphi_D { |}_{x_3 = \pm L/2}=0,
\ee
and Neumann BC:
\be
\label{eq2}
{\partial}\varphi_N/{\partial x_3} { |}_{x_3 = \pm L/2}=0.
\ee
To study the dynamical  effects arising in the translationally
non-invariant case we deal with, it is convenient  
to start with the written in coordinate representation Schwinger-Dyson equation 
(unrenormalized)
for the full 
propagator $G(x,y)\equiv T \langle \phi(x)\phi(y) \rangle$ 
$$
-\partial_x^2\,  G(x,y)= \frac{\lambda}{3!} T \langle \varphi^3(x)  
\varphi(y)\rangle +i\delta(x-y),
$$
which in the leading order in $\lambda$, by virtue of the Wick theorem,
is rewritten as
\be
\label{eq3}
-\partial_x^2\, G_{D,N}=\frac{\lambda}{2}{\cal D}_{D,N}(x,x) G_{D,N}(x,y)
 +i\delta(x-y), %+O(\lambda^2), 
\ee
where ${\cal D}_{D,N}(x,y)$ are the propagators satisfying the free equation
$-\partial_x^2\, D_{D,N}(x,y) =  i\delta(x-y)$, and the Dirichlet
${\cal D}_D{|}_{x_3=\pm L/2}=0$, or  Neumann   
$\partial {\cal D}_N/\partial x_3{|}_{x_3=\pm L/2}=0$ BC, respectively. 
These propagators  are found by the method of mirror images and
in the nontrivial region inside the gap we are interested in, the 
result reads \cite{7}   
%\begin{widetext}
\be
\label{eq4}
{\cal D}_{D, N}(x,y)=-\frac{1}{4\pi^2}\sum_{n=-\infty}^{\infty}
\frac{(\mp 1)^n}{\left(\tilde{x}-\tilde{y}\right)^2
-\left(x_3-(-1)^n y_3-nL \right)^2 }  \\
=-\frac{1}{4\pi^2}
\sum_{n=-\infty}^{\infty}\left[\frac{1}{\left(\tilde{x}-\tilde{y}\right)^2
-\left(x_3- y_3-2nL \right)^2} 
\mp \frac{1}{\left(\tilde{x}-\tilde{y}\right)^2
-\left(x_3+ y_3-(2n-1)L \right)^2} \right],
\ee
%\end{widetext}
where ${\hat x}^2 \equiv x_0^2-x_1^2-x_2^2$. Introducing the quantity 
\be
\label{eq6}
\mu^2_{D,N}(\vec x)=\frac{\lambda}{2}\lim_{x\rightarrow y}\tilde{\cal D}(x,y)
+O(\lambda^2),
\ee
where
\be
\label{eq7}
\tilde {\cal D} (x,y)\equiv {\cal D}(x,y)-{\cal D}_0(x-y), 
\ee
and ${\cal D}_0(x)\equiv -(4\pi^{2})^{-1}[x_0^2-{\vec x}^2]^{-1}$,
one gets from (\ref{eq3}) the equation
\be 
\label{eq8}
\left(-\partial_x^2 -\mu^2_{D,N}(x_3)\right) G_{D,N}(x,y)= i\delta(x-y),
\ee
where now all quantities are renormalized\footnote{ The subtraction 
recipe (\ref{eq7}) means that only nontrivial $L-$dependent part of 
$\mu^2$ survives after renormalization (just as in the case 
of unrestricted space
at finite temperature, where only $T-$dependent part of the dynamical
mass survives after renormalization \cite{3}). It is obvious that another subtraction
schemes can produce in addition only some trivial $L-$independent constants
which always can be absorbed into the infinite mass renormalization constant.
} 
in the leading order\footnote{Notice also that in both  finite quantization region (compactified space dimensions) 
and finite temperature (compactified temporal dimension) cases the renormalization procedure should
be standard (the same as in the infinite space and/or at zero temperature).
Being clear qualitatively, it was also shown for a nonzero temperature \cite{3} by the explicit
calculations beyond the one-loop approximation.
}. 

From (\ref{eq8}) one can see that $x-$dependent quantity $\mu$ 
can be considered an external  ``mass field'' which  acquires the sense of mass 
in its traditional 
(so that 
$E=\sqrt{{\vec p}^2+{\mu}^2}$)
understanding only when weakly (adiabatically) depends on the coordinate.
So, we will call this quantity ``mass gap'' by the analogy with the condense 
matter physics\footnote{Indeed, the same situation
occurs, in particular,
in the second kind superconductors with the chaotic 
distributed additions in the condense matter physics. There  
the energetic gap $\Delta$ generally depends on coordinates. However,   
in the translationally invariant limiting case of homogeneous superconductor,
$\Delta$ determines the dispersion
law
of the elementary excitation and acquires the sense of the quasi-particle mass. 
}.

We first consider the {\it zero temperature} case. 
Using Eqs. (5-7),  the formulas
$
\sum\nolimits_{1}^{\infty} n^{-2}=
\zeta(2)={\pi^2}/{6}$ and,
$$
\sum\nolimits_{-\infty}^{\infty} (n+a)^{-2}
=-\pi\,{d}\,\cot (\pi a)/{da}  ={\pi^2}/{\sin^2(\pi a)},\nonumber 
$$
one easily gets
\be
\label{eq9}
{\mu^2}_{D, N}=\frac{\lambda}{32L^2}\left[\frac{1}{3}
\mp\csc^2\left(\frac{\pi}{L}\left(x_3+\frac{L}{2}\right)\right)\right].
\ee

Let us analyze Eq. (\ref{eq9}). First of all, one can see 
that there are two contributions to $\mu^2$ -- translationally 
invariant and $x_3-$dependent, respectively. The 
translationally invariant contribution
$\lambda /96 L^2$ 
is the same for Dirichlet and Neumann BC
and comes from 
the translationally invariant part
of the propagators ${\cal D}_{D,N}$ (first term in (5).   
Notice that this contribution also can be obtained 
from the well known result of QFT at finite temperature \cite{3}
\be
\label{eq10}
m^2_T =\lambda T^2 /24 
\ee
with the substitution $T\to 1/2L$ 
using the analogy  \cite{10} (similarly to the case of periodic BC \cite{11}) of 
the eigen-frequency spectrum $w_n = \pi n/L$  
for the Dirichlet and Neumann BC
with the finite temperature spectrum $w_n = 2\pi n T$.
However, let us stress 
that in this way one reproduces only a part\footnote{ On the contrary to the periodic BC 
over $x_3$ 
which does not violate translational
invariance. The respective propagator is just the first term in (5) 
(with the replacement $2L\to L$), and, being rewritten to 
Euclidean space, it coincides with the finite temperature propagator with
the replacement $T \to 1/L$. As a consequence, the trick \cite{11} with the substitution $T\to 1/L$
in (\ref{eq10}) just gives the correct $\mu^2$ value for the periodic BC.
} 
of the full $\mu^2$ value 
loosing the $x-$dependent contributions. Moreover, it is seen from (\ref{eq9})
that namely these contributions always dominate 
(i.e., are the biggest at any $x_3$ values). The later leads 
to the crucial difference between Dirichlet and Neumann BC,   
and this is of great importance for what follows: 
while in the case of Neumann BC
mass gap square is always positive $\mu_N^2>0$, 
in the case of Dirichlet BC the mass gap square
is always negative: 
$
\mu_D^2<0,
$
and we will discuss this possibility later.

Looking at Eq. (\ref{eq9}) one can 
notice that the expressions for $\mu^2_{D, N}$ are
divergent on the gap boundaries $\pm L/2$. These divergences 
are not surprising since the propagator (\ref{eq4}) besides of the usual, subtracted by (\ref{eq7}), 
ultraviolet 
singularity $(x-y)^{-2}{|}_{x\to y}$  
 contains also 
 divergent as $x\rightarrow y$  on the boundaries contributions 
corresponding  to $n=\pm 1$ in the sum. 
Such so called ``surface divergences'' are well known \cite{4} from 
the calculation of the Casimir energy density with the boundary conditions 
like (\ref{eq1}) and (\ref{eq2}). It is known that these singularities arise 
because the boundary conditions like (\ref{eq1}) and (\ref{eq2}) are too idealized
approximations to the real ones, and, to avoid the surface divergences
one should deal with more realistic smooth boundary conditions.
However, the transition from idealized sharp boundary
conditions of the total impenetrability to the realistic smooth ones  
cause an enormous complication of all calculations.
Fortunately, there is a possibility to get
reliable results even with the sharp boundary conditions like (\ref{eq1}) and (\ref{eq2}).
Indeed, the experience of the Casimir energy density calculations
show that the application of the smooth boundary conditions instead
of the sharp ones does not influence  the result in the region maximally 
distanced from the boundaries.
So, we will ascribe to result (\ref{eq9}) a physical sense only in such
a region -- in a strip with a small ($\delta/L<<1$) width $\delta$,   
surrounding the central plane $x_3=0$ (see Fig. 1).

On the other hand, this central region has a remarkable property:
since $\partial \mu_{D,N}^2/\partial x_3{\bigl |}_{x_3=0} =0$,
the mass gap there is almost independent from $x_3$ 
$(\mu_{D,N}(\delta)\simeq \mu_{D,N}(0))$ 
and can be considered
a scalar field mass (but not yet as a real mass of 
scalar meson). 

So, in the central $\delta-$region one gets instead of (\ref{eq9}) the 
following expressions\footnote{Certainly, one can reproduce 
the results (\ref{eq11})
at once, putting $x_3=0$ in Eqs. (\ref{eq4}) and (\ref{eq6}) and using that
$\sum\nolimits_{n\ne 0}n^{-2} = \pi^2/3 $ and 
$\sum\nolimits_{n\ne 0}(-1)^n n^{-2} = -\pi^2/6$.
}
for the scalar field masses:       
\be
\label{eq11}
\mu^2_N={\lambda}/24 L^2,\quad
\mu^2_{D}=-{\lambda}/48 L^2.
\ee
Thus, the Neumann and Dirichlet BC differ drastically.
While the Neumann BC leads to the usual
dynamical mass generation of scalar meson:
$m_N=\mu_N=\lambda/24 L^2$, the Dirichlet BC 
leads to the imaginary mass of the scalar field. 
This, as it is well known,
is the signal 
that the spontaneous violation of the ground state symmetry
$\langle\varphi\rangle 
\to\langle\varphi\rangle$
ought to happen, so that after the later,
the scalar meson acquires the real dynamical mass: 
$
m_D^2=2|\mu_D^2|={\lambda}/{24 L^2}.
$
It is of interest that the real meson masses $m_D$ and $m_N$ happen 
to be equal to each other.

Let us show that the result $\mu^2_D<0$ is valid also
in the case where {\it all three space dimensions are compactified}
with the Dirichlet BC satisfied by the scalar field.
Consider the parallelepiped with the edges $L_1,L_2,L_3$ centered around the coordinate origin.
The respective scalar field propagator submitted to the Dirichlet BC on the plates:
${\cal D}_D(x,y)=0$ on $x_i=\pm L_i/2$, has a form \cite{14}
\be
{\cal D}_D(x,y)=\sum\nolimits_{N}(-1)^{(n_1+n_2+n_3)}
{\cal D}_0(x-y^{(N)}),
\ee
where $N=(n_1,n_2,n_3)$, $y_i^{(N)}=(-1)^{n_i}y_i+n_iL_i$
and ${\cal D}_0$ is the free propagator in the infinite space.
Thus, for the cube ($L_1=L_2=L_3\equiv L$) in the small
$\delta$ - region of the coordinate origin, Eqs. (\ref{eq6}), (\ref{eq7}) give
\be
\label{mdcube}
\mu^2_D{\Bigl |}_{cube}=\frac{\lambda}{4\pi^2L^2}\sum'\nolimits_{n_1,n_2,n_3}\frac{(-1)^{n_1+n_2+n_3}}
{n_1^2+n_2^2+n_3^2},
\ee
where symbol prime denotes that
indexes $n_1,n_2,n_3$  in the sum do not equal to zero simultaneously.
Fortunately, the sum entering to (\ref{mdcube}) is a known from the crystal physics 
Madelung constant and can be found in the Table 4 of Ref. \cite{15}:  $\sum'(-1)^{n_1+n_2+n_3}(n_1^2+n_2^2+n_3^2)^{-1}=d(2s)_{s=1}=-2.51935.$ Thus, one again gets the negative result
for $\mu^2_D$:
\be
\mu^2_D{\bigl |}_{cube}=-0.06382\,\lambda/L^2.
\ee

Returning now the flat gap geometry, let us consider {\it  massless scalar electrodynamics}
with the Lagrangian density 
$$%\be
\label{15}
L=-\frac{1}{4}F^2_{\mu\nu}+
\frac{1}{2}\partial^{\mu}\varphi_a \partial_{\mu}\varphi_a
-\frac{\lambda}{4!}\varphi^4 -e\epsilon_{ab}\partial^{\mu}\varphi_a
\varphi_b A_{\mu} +\frac{1}{2}e^2 \varphi^2 A^2,
$$%\ee
where
$\varphi^2\equiv \varphi_a\varphi_a,\,\varphi^4 \equiv {(\varphi^2)}^2,\,
a=1,2,$
and with Dirichlet BC 
$
\varphi_a{|}_{x_3=\pm L/2}=0 \nonumber 
$
satisfied by the scalar field on the gap boundaries. 
Here we will be interested in the dynamical effects caused by the {\it minimal}
modification of the standard (infinite space) QFT. Thus, within 
the present paper, we do not impose\footnote{Though, it is certainly of interest
to investigate in the future also an influence of the nontrivial BC (like Casimir BC)
imposed on the electromagnetic field.} 
any BC on the electromagnetic field,
so that one again has the only tadpole diagram  (for the both fields $\varphi_1$
and $\varphi_2$) contributing to the 
$L-$dependent scalar field mass gap 
in the one-loop approximation.
Operating just as before, one gets in the central region
(see Fig. 1) 
\be
\label{eq12}
\mu^2_D=
-{\lambda}/{36 L^2}.
\ee 
Thus, instead of the massless scalar QED, one arrives at the Higgs model 
with the ``wrong'' sign at the mass term,  and  this occurs
only due to the dynamical corrections in the boundary presence.  

So, after the realization of the standard Higgs mechanism, 
one leaves with 
the only massive scalar meson  with a mass
$ 
m^2_\varphi=2|\mu_D^2|={\lambda}/{18 L^2}
$
interacting with the massive vector field with a mass
$
m^2_A={e^2}|\mu_D^2|/{\lambda} ={e^2}/{36 L^2}.
$

Let us now {\it include the temperature} $T\equiv 1/\beta$ in consideration
and consider first $\lambda \varphi^4/4!$ theory.
Following the well known imaginary time Matsubara approach, one easily
gets instead of (\ref{eq4}), (\ref{eq6}) the equation
$\mu^2_{D,N}=(\lambda/2)\tilde {\cal D}(x,x)+{\cal O}(\lambda^2)$
with
%\begin{widetext}
\be
\label{eq13}
\tilde {\cal D}_{D, N}(x,y)=(4\pi^2)^{-2}{\sum'\nolimits_{n,m}}
{(\mp 1)^n}[\left(x_4- y_4+m\beta\right)^2
+\left(\tilde{x}-\tilde{y}\right)^2
+\left(x_3-(-1)^n y_3+nL \right)^2 ]^{-1},
\ee
%\end{widetext}
where $\tilde x\equiv (x_1,x_2)$ and the symbol prime denotes that
indexes $n$ and $m$ in the sum do not equal to zero simultaneously.
Then, in the case of Dirichlet BC, near the central 
plane $x_3=0$, one gets
\be
\label{eq14}
\mu^2_D(L,\beta)=({\lambda}/{8\pi^2}){\sum'\nolimits_{n,m}}
{(-1)^n}[{\beta^2 m^2+L^2 n^2}]^{-1}.
\ee   
Using ${\sum\nolimits_n}^{'} (-1)^n n^{-2}=-\pi^2/6 $
and ${\sum\nolimits_n}^{'}  n^{-2}=2\zeta(2)=\pi^2/3 $
one can easily get the asymptotes $\mu^2_D(L,\infty)$
and $\mu^2_D(\infty,\beta)$ corresponding to the zero temperature
and infinite space cases, respectively. In the first case
one arrives at the result (\ref{eq11}) for $\mu_D^2$, and, in the second case one 
obtains the 
result (\ref{eq10}): $m^2_T=\mu_D^2 (\infty,\beta)=\lambda T^2/24$.

Let us now introduce two new evolution variables 
\be
\label{eq15}
\chi_1\equiv \beta^2/L^2, \quad \chi_2\equiv \chi_1^{-1} =L^2T^2.
\ee

Calculating the sums
${\sum'\nolimits_{n,m}}
(-1)^n[\chi_1 m^2+ n^2]^{-1}$
and
${\sum'\nolimits_{n,m}}
(-1)^n[ m^2+ \chi_2 n^2]^{-1}$
one obtains the respective evolution pictures presented
by Fig. 2 (top and bottom pictures, respectively).
\begin{figure}[ht!]
\begin{center}
        \includegraphics[height=6.0cm,width=8.6cm]{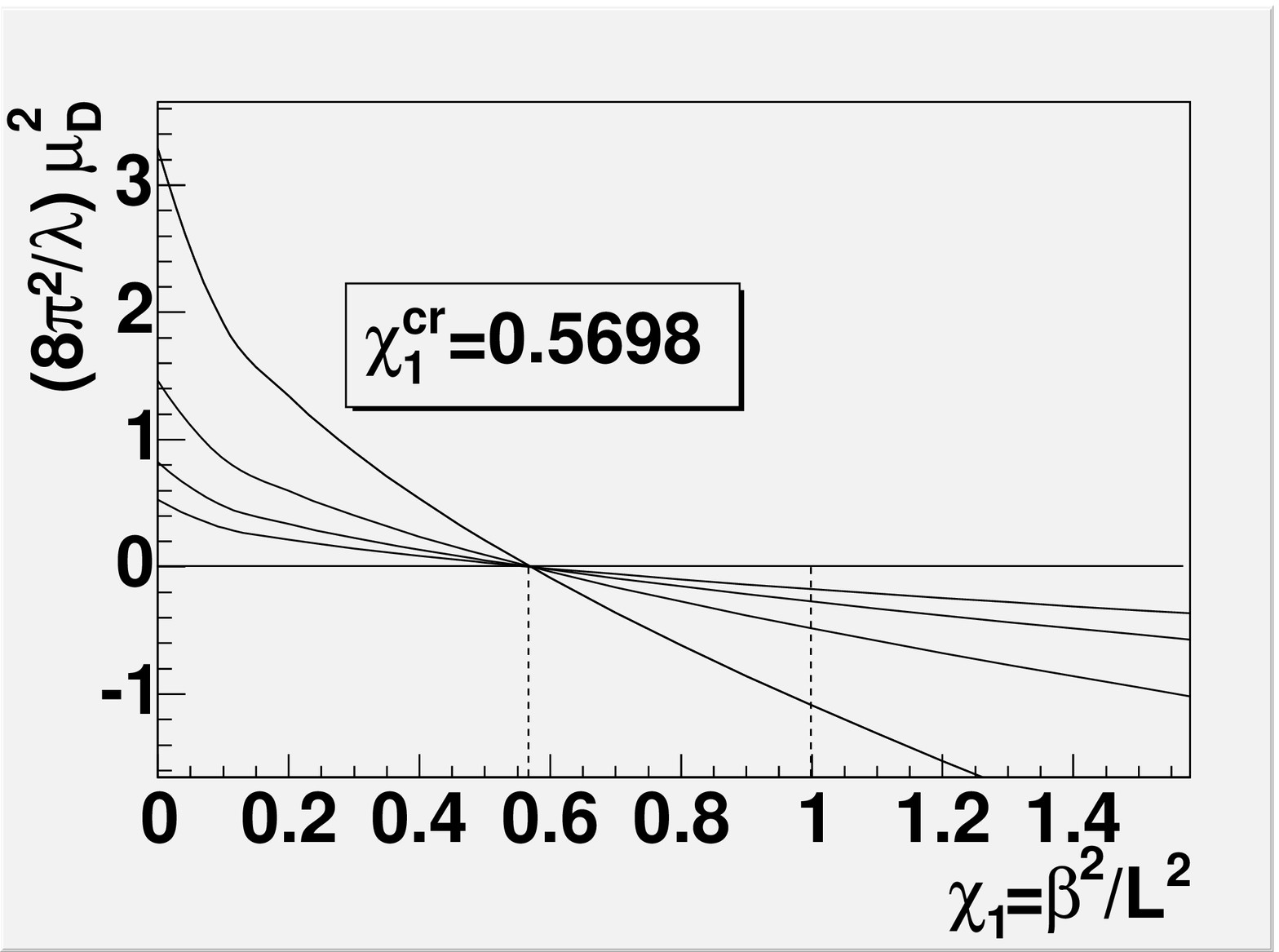}
        \includegraphics[height=6.0cm,width=8.6cm]{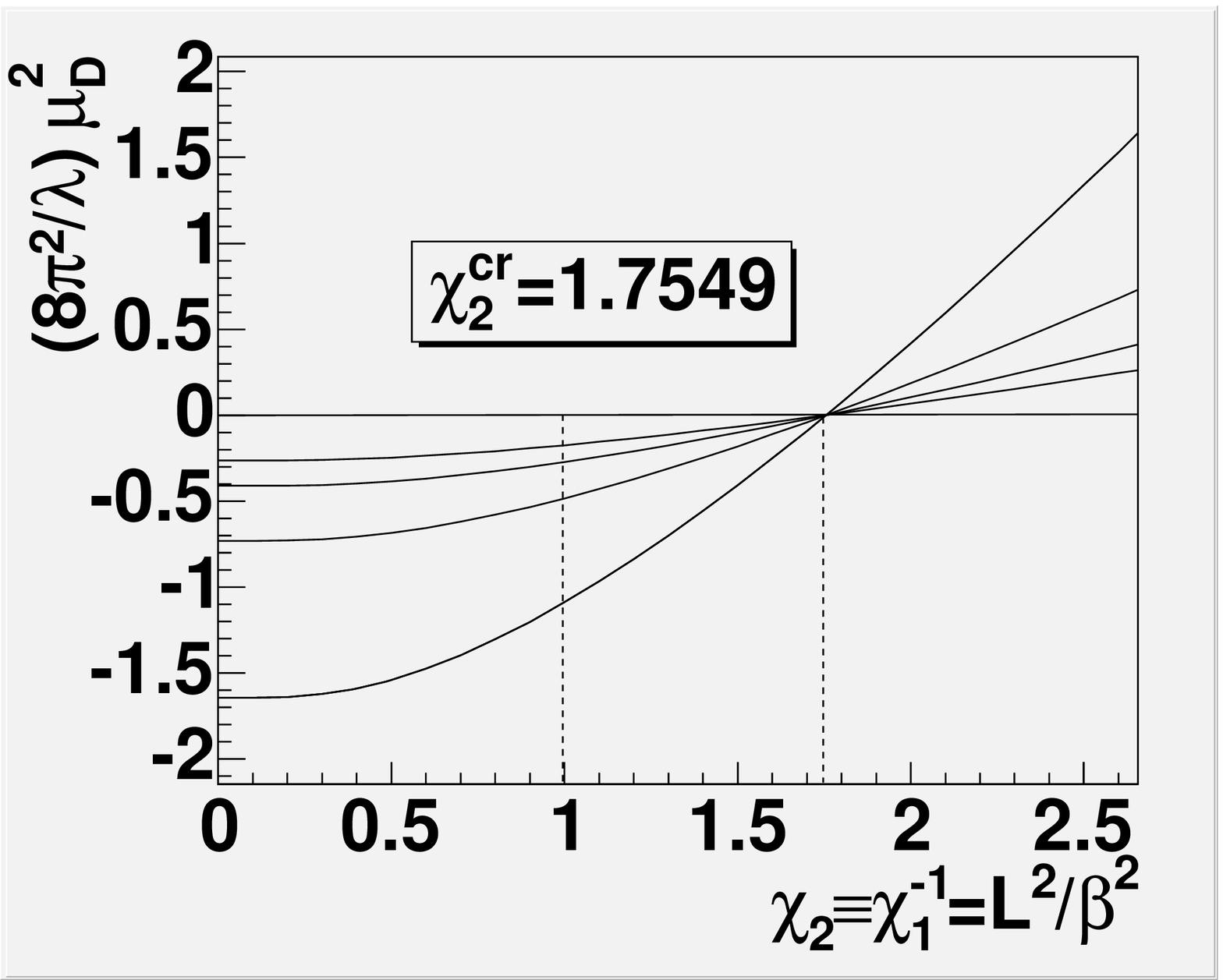}
\end{center}
\caption{$(8\pi^2/\lambda)\mu_D^2$ versus 
$\chi_1$ (top) and $(8\pi^2/\lambda)\mu_D^2$ versus 
$\chi_2\equiv \chi_1^{-1}$ (bottom). The natural system $\hbar=c=k=1$ is used. 
For  each curve on the top picture  $\beta$ is fixed
while $L$ evolves and, otherwise, for the bottom picture.
}
\end{figure}
One can see the critical point $\chi_1^{cr} = 0.5698$ at which $\mu^2_D$
changes the sign, i.e., there occurs the phase transition from the normal to the spontaneously
broken phase. It is of importance that the phase transition
can occur either because of changing the temperature at fixed $L$ (bottom picture)
or because of the gap size $L$ changing at fixed temperature (top picture).  
The asymptotes of $\mu^2_D$ as $\chi_1\rightarrow0$ ($L\rightarrow\infty$ while $\beta$
is fixed) and as $\chi_2\rightarrow0$  ($\beta\rightarrow\infty$ while $L$ is fixed)
are presented by the left edges of the top and bottom pictures, respectively, and are
in agreement with Eqs. (\ref{eq10}), (\ref{eq11}). The results for the symmetric
point $\beta=L$ ($\chi_1=\chi_2=1$) are in agreement with the exactly
calculated \cite{16} sum $\sum'\nolimits_{n,m}(-1)^n[n^2+m^2]^{-1}=(-1/2)\pi\ln 2$.

Let us also stress the {\it essential advantage} of the just considered  
dynamical mechanism  of the spontaneous 
symmetry breaking (restoration). 
In our case, there are no problems with the perturbative calculation of the critical point 
as it occurs at investigation of the spontaneously broken symmetry restoration at critical temperature \cite{3}, 
since
we do not introduce into the Lagrangian the imaginary mass term by hand from the very
beginning, that leads\footnote{Already in the one-loop approximation one arrives \cite{3} at  complex value for the critical temperature (see \cite{3} p. 3324), so that the results become  physically acceptable only in the $\beta\rightarrow0$ limit.
} to the complex value for the critical
point.

It is obvious that in the case of {\it the massless scalar electrodynamics} 
with the Dirichlet BC on the gap plates, 
the evolution pictures are analogous to the presented by Fig. 2 ones 
with the same critical point. The only difference is in the asymptotes.
Namely, $\mu^2_D \to \lambda T^2/18$ as $\chi_1 \to 0$ ($L \to \infty$ while $T$ is fixed)
and $\mu^2_D$ tends to the zero temperature result (\ref{eq12}) as $\chi_2\to 0$ 
($\beta \to \infty$ while $L$ is fixed).
So,  because of the gap size decreasing at fixed temperature 
at $\chi_1^{cr} =0.5698$, there occurs the phase transition: the massless
scalar electrodynamics with the Dirichlet BC satisfied by the scalar fields 
on the gap boundaries
transforms to the Higgs model with the spontaneous symmetry violation.
As a result, at $\chi_1>\chi_1^{cr}$, after the realization of the standard
Higgs mechanism,  one leaves with 
the only massive scalar meson interacting with the massive vector boson. 
 
Thus one can say that  the boundary 
with the respective BC (Dirichlet here)
and  the temperature compete with each other:
while the temperature  always aspires to restore the broken symmetry,
the boundary tends to violate it. 
This competition gives rise to the new type 
of phase transition:
the decreasing in the characteristic size of the quantization region
(the gap size here) and the increasing in the temperature 
tend to transport the system into the spontaneously broken or
into the normal phase.
The system evolves with a combined parameter reflecting the change
in the temperature and in the size
simultaneously. As a result, at the critical value of this parameter
there occurs the phase transition from the normal to the
spontaneously broken phase. In  particular, the usual massless
scalar electrodynamics transforms to the Higgs model. The later,    
as it is well known, is the key model supporting the foundations of main directions  
in the modern physics based on the spontaneous symmetry breaking and the Higgs mechanism.
In particular, these are 
superconductivity theory\footnote{In the effective superconductivity theory the electron pairs (Cooper pairs) are absent outside of the superconductor. 
To provide this "confinement" it is natural to impose the Dirichlet BC  on the wave
function describing (see \cite{19} and references therein) the Cooper pairs.
}   (which is just the nonrelativistic 
variant of the Abelian Higgs model -- see, for example, \cite{19} and references therein) 
in the condense matter physics, and the Weinberg-Salam theory of 
the electroweak interactions in the high energy physics. So, one can hope that the presented here
dynamical phenomena caused by the boundary influence can lead to some new
physical predictions in these important branches of the modern physics.   
%\begin{acknowledgments}
        \vskip 0.3cm
We are grateful to the specialists from the Scientific Center for Applied Research
at JINR G. Emelyanenko and O. Ivanov for the help in performing of numerical calculations.
We also  grateful to N. Kochelev, E. Kuraev and  S. Nedelko  for fruitful discussions.
%\end{acknowledgments}
 
\end{document}